\documentclass[floatfix,useAMS,usenatbib]{mn2e}
\usepackage{graphicx,epsfig}
\usepackage{multirow}
\usepackage{verbatim}
\usepackage[figuresright]{rotating}
\usepackage{txfonts}

\usepackage{graphicx}
\usepackage{graphpap}
\usepackage{epsf}
\usepackage{amssymb}
\usepackage{color}

\def\FIG #1 #2 [#3] #4\par{%
  \begin{figure}\begin{center}%
    \includegraphics*[#3]{#2}%
    \\
    \caption{#4}%
    \label{#1}%
  \end{center}\end{figure}%
}
\def\FIGG #1 #2 #3 [#4] #5\par{%
  \begin{figure*}\begin{center}%
               \includegraphics*[#4]{#2}
               \includegraphics*[#4]{#3}
    \caption{#5}%
    \label{#1}%
   \end{center}\end{figure*}%
}
\def\FIGth #1 #2 #3 #4 [#5] #6\par{%
  \begin{figure*}\begin{center}%
        \includegraphics[#5]{#2} 
        \includegraphics[#5]{#3}
        \includegraphics[#5]{#4}
        \caption{#6}
        \label{#1}
   \end{center}\end{figure*}
}
\def\FIGtha #1 #2 #3 #4 [#5] #6\par{%
  \begin{figure*}\begin{center}%
        \includegraphics[trim=0 0 55 10,clip=true,width=0.3663\hsize,angle=-0]{#2} 
        \includegraphics[trim=80 0 55 10,clip=true,width=0.31\hsize,angle=-0]{#3}
        \includegraphics[trim=80 0 55 10,clip=true,width=0.31\hsize,angle=-0]{#4}
        \caption{#6}
        \label{#1}
   \end{center}\end{figure*}
}
\def\FIGthx #1 #2 #3 #4 [#5] #6\par{%
  \begin{figure*}\begin{center}%
        \includegraphics[trim=10 62 40 30,clip=true,width=0.9\hsize,angle=-0]{#2} 
        \includegraphics[trim=10 62 40 30,clip=true,width=0.9\hsize,angle=-0]{#3}
        \includegraphics[trim=10 0 40 30,clip=true,width=0.9\hsize,angle=-0]{#4}
        \caption{#6}
        \label{#1}
   \end{center}\end{figure*}
}
\def\FIGfo #1 #2 #3 #4 #5 [#6] #7\par{%
  \begin{figure*}\begin{center}%
        \includegraphics[#6]{#2}
        \includegraphics[#6]{#3} \\
        \includegraphics[#6]{#4}
        \includegraphics[#6]{#5}
        \caption{#7}
        \label{#1}
   \end{center}\end{figure*}
}
\def\FIGfi #1 #2 #3 #4 #5 #6 [#7] #8\par{%
  \begin{figure*}[ht]\begin{center}%
        \includegraphics[#7]{#2}
        \includegraphics[#7]{#3} 
        \includegraphics[#7]{#4} \\
        \includegraphics[#7]{#5}
        \includegraphics[#7]{#6}
        \caption{#8}
        \label{#1}
   \end{center}\end{figure*}
}
\def\FIGsi #1 #2 #3 #4 #5 #6 #7 [#8] #9\par{%
  \begin{figure*}[ht]\begin{center}%
        \includegraphics[#8]{#2}
        \includegraphics[#8]{#3}\\
        \includegraphics[#8]{#4}
        \includegraphics[#8]{#5}\\
        \includegraphics[#8]{#6}
        \includegraphics[#8]{#7}
        \caption{#9}
        \label{#1}
    \end{center}\end{figure*}
}
\def\FIGei #1 #2 #3 #4 #5 #6 #7 #8 [#9] {%
        \Figrow #1 #2  [#9]
        \Figrow #3 #4 [#9]
        \Figrow #5 #6 [#9]
        \Figrow #7 #8 [#9]
}

\def\Figrow #1 #2 [#3]{%
        \includegraphics[#3]{#1}
        \includegraphics[#3]{#2}\\
}

\def\rfig#1{Figure \ref{#1}}

\def\msun{M_\odot}

\newcommand{\mt}[1]{\mathrm{#1}}
\newcommand{\sect}[1]{Section~\ref{#1}}
 
\def\lvm{\leavevmode\hbox to\parindent{\hfill}}

\def\BE{\begin{equation}}
\def\EE{\end{equation}}
\def\BA{\begin{array}}
\def\EA{\end{array}}
\def\BAN{\begin{eqnarray*}}
\def\EAN{\end{eqnarray*}}

\def\la{\mathrel{\mathpalette\fun <}}
\def\ga{\mathrel{\mathpalette\fun >}}
\def\fun#1#2{\lower3.6pt\vbox{\baselineskip0pt\lineskip.9pt
\ialign{$\mathsurround=0pt#1\hfil##\hfil$\crcr#2\crcr\sim\crcr}}}

\def\gccm{\mathrm{g\,cm}^{-3}}

\def\msun{M_\odot}
\def\Msun{$\msun$}

\def\e#1{$\times 10^{#1}$ }
\def\ee#1{$10^{#1}$ }

\def\ltsima{$\; \buildrel < \over \sim \;$}
\def\ltsim{\lower.5ex\hbox{\ltsima}}
\def\gtsima{$\; \buildrel > \over \sim \;$}
\def\gtsim{\lower.5ex\hbox{\gtsima}}

\def\nism{n_0}

\def\Pcr{P_\mt{CR}}

\def\qcr{q_\mt{CR}}

\def\Syng{SNR~0509-67.5}
\def\Sold{SNR~0519-69.0}
\def\cc{\,\mathrm{cm}^{-3}}
\newcommand{\ions}[2]{#1~{\sc #2}}

\usepackage{times}
\usepackage{txfonts}
\usepackage{natbib}
\bibpunct{(}{)}{,}{a}{}{,}

\begin{document}

\title[Oxygen in thermonuclear SNRs]{Oxygen emission in remnants of
  thermonuclear supernovae as a probe for their progenitor system}

\author[D. Kosenko, W. Hillebrandt, M. Kromer, S.I. Blinnikov,  
  R. Pakmor and J.S. Kaastra ]
  {D. Kosenko$^{1}$\thanks{E-mail:daria.kosenko@gmail.com}, 
  W. Hillebrandt$^{2}$, M. Kromer$^{3,2}$, S.I. Blinnikov$^{4,5,6}$, 
  R. Pakmor$^{7,2}$
  \newauthor
  and J.S. Kaastra$^{8,9}$\\ 
  $^{1}$Sternberg Astronomical Institute (MSU), 119992, 
    Universitetskij pr.13., Moscow, Russia \\
  $^{2}$Max-Planck-Institut f\"ur Astrophysik, 
    Karl-Schwarzschild-Stra{\ss}e 1, D-85748 Garching, Germany\\
  $^{3}$The Oskar Klein Centre \& Department of Astronomy, 
    Stockholm University, AlbaNova, SE-106 91 Stockholm, Sweden\\
  $^{4}$Kurchatov Institute for Theoretical and Experimental Physics, 
    Bolshaya Cheremushkinskaya 25, 117218 Moscow, Russia \\
  $^{5}$Kavli Institute for the Physics and Mathematics of the Universe (WPI),  
  5-1-5 Kashiwanoha, Kashiwa, Chiba 277-8583, Japan \\
  $^{6}$VNIIA, Moscow 127055, Russia\\  
  $^{7}$Heidelberger Institut f\"{u}r Theoretische Studien, 
    Schloss-Wolfs\-brunnen\-weg 35, 
    69118 Heidelberg, Germany\\
  $^{8}$SRON Netherlands Institute for Space Research, 
    Sorbonnelaan 2, 3584 CA Utrecht, The Netherlands \\
  $^{9}$Leiden Observatory, Leiden University, P.O. Box 9513, 2300 RA Leiden, The
Netherlands
  }

\maketitle

\begin{abstract} 

  Recent progress in numerical simulations of thermonuclear supernova
  explosions brings up a unique opportunity in studying the
  progenitors of Type Ia supernovae. Coupling state-of-the-art
  explosion models with detailed hydrodynamical simulations of the
  supernova remnant evolution and the most up-to-date atomic data for
  X-ray emission calculations makes it possible to create realistic
  synthetic X-ray spectra for the supernova remnant phase. Comparing
  such spectra with high quality observations of supernova remnants
  could allow to constrain the explosion mechanism and the progenitor
  of the supernova. The present study focuses in particular on the oxygen emission
  line properties in young supernova remnants, since different
  explosion scenarios predict a different amount and distribution of
  this element.  Analysis of  the soft X-ray spectra from supernova remnants in the Large
  Magellanic Cloud and confrontation with remnant models for different
  explosion scenarios suggests that \Syng\ could originate from a
  delayed detonation explosion and \Sold\ from an oxygen-rich merger. 

\end{abstract}



\section{Introduction}

There is general consensus that SNe Ia are exploding white dwarfs
(WDs), composed of carbon and oxygen, where the explosion is triggered
by mass transfer from a companion star \citep[e.g.][]{hillebrandt00}. Traditionally, two classes of
potential binary progenitor systems are distinguished: The
single-degenerate progenitor channel (SD), in which the companion of
the WD is a normal star \citep{whelan73}, and the double-degenerate channel (DD) with
two WDs interacting and merging \citep{iben84}. At present it is unclear whether one
of these possibilities is exclusively realized in nature or whether
both contribute to the class of SNe Ia. In the past, it was
believed that in the SD scenario the exploding WD has to be
close to the Chandrasekhar-mass limit to ignite explosive carbon
burning, first as a subsonic combustion wave (deflagration) with a
likely transition to a detonation after it has expanded
to somewhat lower density, in order to be in agreement with the
observed lightcurves and spectra. The major disadvantage of this
scenario was too low a number of events predicted by binary population
synthesis. The DD scenario, on the other hand, had the
disadvantage that theory predicted that the process of merging would
lead to an accretion-induced collapse rather than a thermonuclear
disruption \citep[see][for a review]{hillebrandt00}.

More recently, supported by increasing evidence from observations, the
picture has emerged that SNe Ia are a more diverse class of
objects \citep{li11}. The SD channel now includes WDs that explode with a mass
significantly below the Chandrasekhar mass, the explosion being
triggered by the detonation of a layer of helium on top, accreted from
the companion, the so-called 'sub-Chandrasekhar double-detonation'
scenario \citep{livne90,fink07,moll13}. In fact, this scenario is not limited to the SD channel. It
may happen that during the process of merging helium from the outer
layers of the companion WD is accreted onto the primary and burns
explosively there \citep{pakmor13}. In addition, the DD channel may include binaries
with a He WD as secondary companion instead \citep[see][for a recent review]{hillebrandt13}.

For our present study it is important that the various explosion
scenarios yield element abundances and distributions which can be
rather different and can, in principle, be distinguished by
observations. For instance, Chandrasekhar-mass delayed-detonation
models and the sub-Chandrasekhar ones predict generic differences for
the iron-group elements because, in the first scenario, subsonic
turbulent nuclear burning at high density goes together with
convective mixing \citep{seitenzahl13}, whereas in the second scenario the density at
ignition, set by the WD's mass, determines the final abundances and
their distribution, and there is no mixing \citep{sim10, shigeyama92}. Some DD mergers, on the
other hand side, may have iron-group abundances very similar to the SD
sub-Chandrasekhar explosions if the secondary WD is left behind
unburned.

As far as lighter elements, such as oxygen, are concerned significant
differences exist between the various scenarios as well. In 
DD mergers with a CO WD companion there may be a lot of unburned
oxygen in the system, by default a generic feature of models in which
the secondary is disrupted but not completely burned \citep{pakmor12}. In contrast, delayed-detonation
Chandrasekhar-mass models have little oxygen, but at high velocity \citep{seitenzahl13}, as
have accreting sub-Chandrasekhar models \citep{fink10}. In
pure-deflagration Chandrasekhar-mass models, finally, due to mixing,
low-velocity oxygen could also be present \citep{fink14}. It is the main goal of this
paper to investigate if such differences may leave signatures in the
X-ray emission during the supernova's remnant phase.

In a supernova remnant (SNR) when the reverse shock propagates inwards
into the SN ejecta, the heated ejecta start to produce thermal X-ray
emission, whose properties are determined by the chemical composition
of the plasma \citep[][for a review]{vink12}. We perform numerical hydrodynamical simulations of an
SNR evolution for different thermonuclear explosion scenarios.
Analysis of the synthetic soft X-ray spectra built from these
hydrodynamical models shows that the properties of the respective
oxygen emission lines systematically vary from model to model. 

To assess the significance of these variations we confront the
numerical models with observations of young SNRs located in the Large
Magellanic Cloud (LMC), \Syng\ and \Sold. To estimate the
oxygen contribution to the emission, we use data from the reflective
grating spectrometer (RGS) on board the \emph{XMM Newton} observatory.
We consider these objects as good case studies as the available high
resolution soft X-ray spectra clearly show the presence of bright
oxygen emission lines \citep{kosenko08, kosenko10}. Moreover, these objects are fairly well
studied, so that estimates of their dynamical properties (such as size, 
age and ISM density) are available in the literature.

The structure of the paper is as follows. We outline the explosion
models used in this study in \sect{models} and describe our method in
\sect{method}. Results from our simulations and the corresponding
synthetic X-ray spectra are discussed in \sect{simulations}. A
detailed description of two LMC remnants and an analysis of the oxygen
lines in their spectra are presented in \sect{oxygen}. A comparison of
our models with the observed data is given in \sect{lmcdata}. Finally,
we discuss our results in \sect{results} and conclude in
\sect{conclusion}.

\section{Explosion models}
\label{models}

For the present study we have picked three hydrodynamical models which
may be considered as being 'typical' for some of the explosion
scenarios introduced in Sect. 1, and for explaining 'normal' SNe
Ia. The models were chosen to produce about $0.6\,M_\odot$ of
$^{56}$Ni in all cases such that the predicted luminosity at peak
would agree with 'normal' SNe Ia.

The first model (``subch'', $1.1\times$\ee{51}erg) is from the class 'sub-Chandrasekhar double
detonations' as described in \citet{fink10} and \citet{sim10}.
In the work of \citet{fink10} the core detonation is caused by the
detonation of a thin He shell of a few $10^{-2}\,M_\odot$. In
contrast, in the \citet{sim10} work the CO core detonation is an
assumption and the effect of the He layer on light-curve and spectra
is ignored, the goal being to study the general behaviour of such
models rather than detailed spectra. The particular model we use here
is not described in those papers but was chosen to match the
radioactive Ni mass of the other two models discussed below. It
consists of a bare $1.1\,M_\odot$ CO WD with an artificially ignited
central detonation. After explosion, the simulation yields
$0.615\,M_\odot$ of $^{56}$Ni and the mass of oxygen is
$0.07\,M_\odot$, with velocities from about 14,000 to more than 25,000
km/s. Synthetic light-curve and spectra resemble normal SNe Ia rather
well.

The second model (``deldet'', $1.4\times$\ee{51}erg) is a 'classical' delayed-detonation model \citep{roepke12,seitenzahl13}. 
In this model, a Chandrasekhar-mass
WD was set up in hydrostatic equilibrium with a central density of
$2.9\times10^9\,\gccm$ and an electron fraction of $Y_e=0.498864$,
corresponding to solar metallicity. An initial deflagration was ignited
in $100$ sparks placed randomly in a Gaussian distribution within a
radius of $150\,\mathrm{km}$ from the WD's center in order to get
approximately $0.6\,M_\odot$ of $^{56}$Ni. 
After an initial deflagration phase a detonation was triggered based on the
strength of turbulent velocity fluctuations at the flame front \citep[for details
on the treatment of the deflagration-to-detonation transition see][]{Ciaraldi-Schoolmann13}.
The evolution was followed to a time of
$100\,\mathrm{s}$ after ignition, by which homologous expansion of the
ejecta was reached to a good approximation. This model fits the data of
SN 2011fe in M101 reasonably well \citep{roepke12}. The mass of
oxygen predicted by the model is $0.101\,M_\odot$, slightly above the
sub-Chandrasekhar model presented here. Typical oxygen velocities
are in the range from 10,000 to 20,000 km/s.

The third model (``merger'', $1.6\times$\ee{51}erg) we will use for our comparison with X-ray data is the
inspiral, merger, and explosion of two CO WDs with $1.1\,M_\odot$ and
$0.9\,M_\odot$, respectively. Details of the corresponding simulations
are given by \citet{pakmor12}. 
In this simulation, the inspiral
and merger phases were followed with a version of the SPH code
\textsc{gadget} \citep{springel05} 
and the subsequent thermonuclear
detonation was modeled with techniques similar to those employed in the
delayed-detonation model described before. The question of whether a
detonation triggers at the interface between the two merging stars is
controversial. In our simulations we assumed a detonation to trigger
when in one location the temperature exceeded $2.5\times
10^{9}\,\mathrm{K}$ in material of $\rho\approx 2\times
10^{6}\,\gccm$ \citep{Seitenzahl09}. Again, the evolution was followed up to
$100\,\mathrm{s}$ and the composition of the ejecta was determined in
a post-processing step. Also this model was used for a comparison
with SN 2011fe by \citet{roepke12}. 
Here the oxygen mass in the
ejecta is considerably higher than in the two other models, namely
$0.492\,M_\odot$ and the abundance distribution, in particular that of
the iron-group elements is far less symmetric. Typical oxygen
velocities range from a few 1000 to 20,000 km/s.  

For all three models the element abundances and their
distribution differ (\rfig{abunds}) 
which will become important when we discuss their X-ray spectra.

\FIGtha abunds {subch_200_ISM30_v_abn} {deldet_200_ISM30_v_abn}
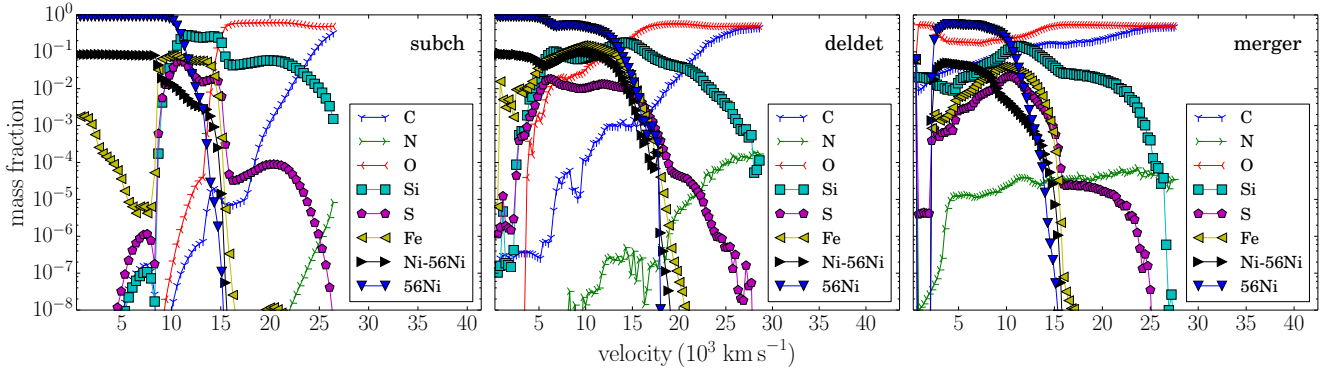
{merger_200_ISM30_v_abn} [] Spherically averaged radial abundance profiles of the most prominent elements as functions of ejecta velocity. The models age is 10 yrs. 


\section{Method}
\label{method}

For our hydrodynamical (HD) simulations of the SNR evolution we apply
the {\sc supremna} code \citep{blinnikov98,sorokina04, kosenko06}.
{\sc supremna} uses an implicit Lagrangian formulation and takes into
account the most important relevant physical processes, such as
electron thermal conduction and a self-consistent calculation of the
time-dependent ionization of the shocked plasma. The contribution of
relativistic cosmic-ray particles to the dynamics is treated in a
two-fluid approximation \citep{kosenko11}. Since {\sc supremna}
assumes spherically-symmetric ejecta, we use spherically averaged
versions of the multi-dimensional explosion models described above, 
keeping in mind that there could be complications for the ``merger'' model.

The HD package is coupled with the most up-to-date atomic data from
the {\sc spex} spectral fitting software \citep[][]{kaastra96} to
obtain synthetic X-ray spectra from the modeled SNR. The call of the
{\sc spex} routine is executed for each cell of the HD grid. The total
spectra are obtained by integrating over the entire model.

\section{Numerical simulations}
\label{simulations}

The SNR evolution was modeled for each of the three explosion
scenarios. We start the simulations at a time of $\sim3$ years after
the explosion (the explosion models were expanded accordingly with the unstable nickel decayed to iron). The
remnants expand through a uniform interstellar medium (ISM) with a
temperature of \ee{4}K and various densities $\nism$. We set the ratio of
electron to ion temperature to $0.3$. Electron thermal conduction was
suppressed \citep{sorokina04} and diffusive shock acceleration (DSA)
processes were neglected. However, we run one model with a moderate DSA efficiency, 
in order to estimate its impact on the final results.

To illustrate how the composition of the models influence the emission from remnants, synthetic X-ray spectra in the range of $(0.45-0.7)$ keV at a remnant's age of $\sim430$ yr are presented in \rfig{xsprgs}. The spectra are shown for different values of  $\nism$ (marked in the middle column). The lines of \ions{O}{vii} at 0.57 keV, \ions{O}{viii} at 0.65 keV, and \ions{N}{vii} and \ions{N}{vi} at 0.5 keV (marked in the middle row) are visible for all models and ISM densities. 

The spectra in \rfig{xsprgs} reveal that in the ``subch'' and ``deldet''  models most of the carbon is depleted, while the ``merger'' is  rich in the light elements.  In the ``subch'' and ``deldet'' models the ejecta \ions{C}{vi} contribution at 0.46 keV is absent, and only the \ions{S}{xiii} complex at 0.47 keV is noticeable for $\nism\gtrsim1.0\cc$. In the ``merger'' spectra the sulfur contribution cannot be distinguished as it mixes with the relatively bright carbon line seen as an emission excess at $0.45-0.49\;\mt{keV}$. 

\FIGthx xsprgs {xsp_ISM03__t0440_e04-07_spx} {xsp_ISM10__t0429_e04-07_spx} {xsp_ISM30__t0426_e04-07_spx} 
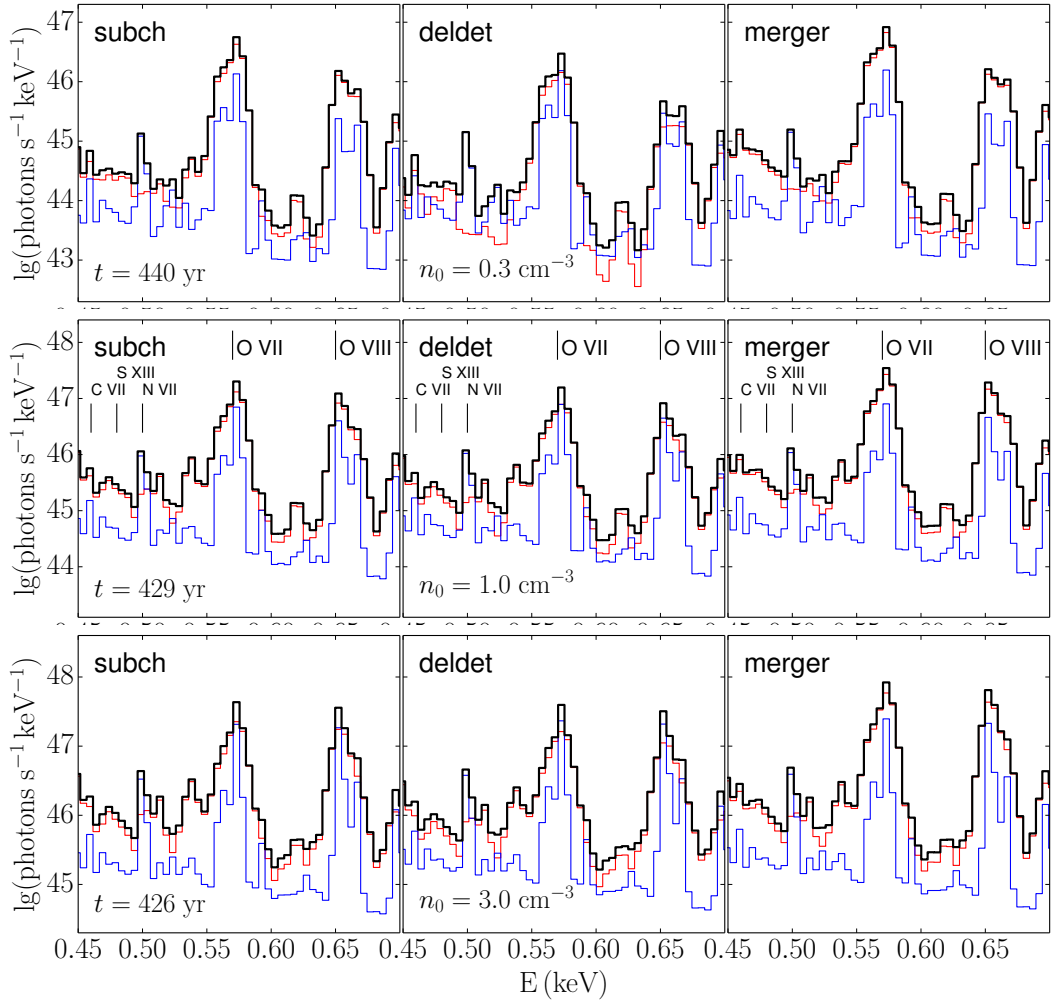
[] Synthetic X-ray spectra for different ambient medium densities. \ions{S}{xv} is at 0.45 keV, \ions{C}{vi} line is at 0.46 keV, \ions{S}{xiii} is at 0.46-0.47 keV, \ions{N}{vii} line is at 0.5 keV,  \ions{O}{vii} is at 0.57 keV, and \ions{O}{viii} is at 0.65 keV (marked in the middle row). Blue thin lines are for the shocked ISM, thin red for the shocked ejecta, thick black line is the total emission. The exact time of the model is in the left column, the ISM density is indicated in the middle column. 

\section{Analysis of the oxygen emission lines for the various
  explosions}
\label{oxygen}

To estimate the line brightness, we consider the flux ($F_X$) of a
line and its equivalent width EW~$= \int (F_X - F_0)\,dE/\int
F_0\,dE$, where $F_0$ is the continuum flux at the line centroid.
Furthermore, we calculate ratios of the oxygen emission line
characteristics relative to those of nitrogen. 
Although the ``merger'' model yields noticeable amount of nitrogen (mass fraction of $\sim3\times10^{-5}$), it is still negligible compared to the solar abundance of  $\sim10^{-3}$ and to the nitrogen fraction in the LMC, which we assume is approximately factor of three less than that of the Milky Way \citep[e.g.][]{Welty99,Piatti13}.
By relating the ejecta emission to the emission from the shocked ISM,
we partially eliminate the effects of the unknown properties of the
underlying continuum.

The EW is a measure of the total amount of the emitting particles. It
takes into account all the particles with different velocities, thus
accounting for the effects of the Doppler broadening. However, the EW
depends on the continuum level around the line, which includes thermal
and non-thermal components and depends on unknown plasma parameters
(ion and electron temperatures, density, filling factor, etc.). In addition, our
1D hydrodynamical method cannot predict the X-ray continuum with
sufficient accuracy.

Using the ratios of the line fluxes alleviates the influence of the
continuum but does not account for the Doppler effect. However, we
expect the line velocity broadening of the shocked oxygen in the
ejecta and of the shocked nitrogen in the ISM to be correlated.
A time evolution of the dimensionless EWs of
\ions{O}{vii}/\ions{N}{vii} (solid) and \ions{O}{viii}/\ions{N}{vii}
(dashed) for our models and different ISM densities is presented in
\rfig{o_ew_evol}.

\FIGth o_ew_evol {ISM03_00_ratio_EW_evol} {ISM10_00_ratio_EW_evol}
{ISM30_00_ratio_EW_evol} [trim=10 0 40
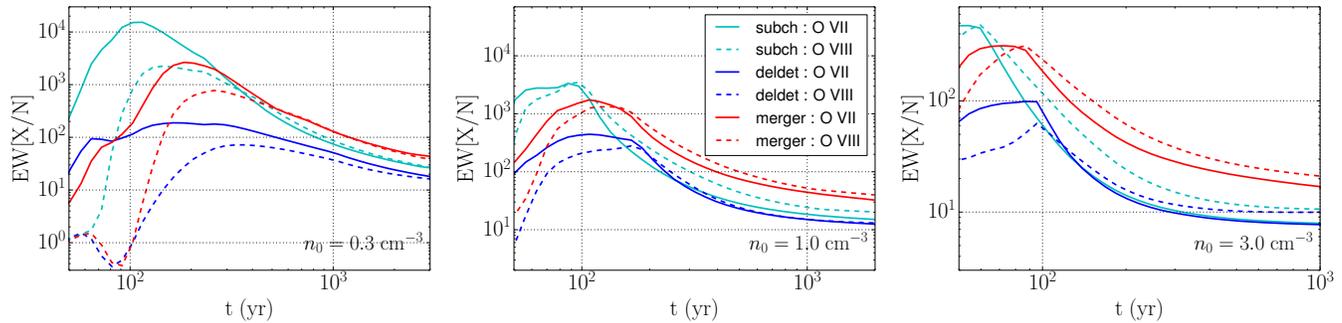
30,clip=true,width=0.33\hsize,angle=0] Evolution of the equivalent
width ratios of \ions{O}{vii}/\ions{N}{vii} (solid) and
\ions{O}{viii}/\ions{N}{vii} (dashed) for different ambient densities
($\nism=0.3,\,1.0,\,3.0\cc$ from left to right, respectively).

The time when the profiles reach their maximum is determined by the
energy of the respective explosion. The amplitude of the EW ratios
depends on the amount of oxygen. In remnants younger than 100-200\,yr
the ISM nitrogen line is rather weak, which explains the high values
of the EW ratios at this stage. At an age of several hundred years the
nitrogen line becomes pronounced and the respective EW ratios drop.

At low ISM densities and early times of $\lesssim 100$~yr (the left plot of \rfig{o_ew_evol}) the ``deldet'' and ``merger'' profiles both produce weak oxygen emission relative to nitrogen. Different oxygen abundance is balanced by the different explosion energy, so that the normalised oxygen emissivity is comparable. However, later (e.g. starting from 100 yr for the left plot with $\nism=0.3\cc$) the ``merger'' scenario yields higher oxygen emmisivity. 

For the high density models the ``merger'' profiles clearly stand out compared to the other two cases.
The profiles for the ``deldet'' model are lower, as this scenario has almost as much oxygen
as the ``subch'' model while it is almost as energetic as the
``merger'' model. High velocity outer layers of the ejecta lead to the
expansion of the forward shock (FS) as fast as in the ``merger'' model
(see the dynamics in the left column plots of \rfig{fig:o_rad})
involving more nitrogen from the ambient medium, which reduces the
respective normalized EW. Thus, relatively weak oxygen lines are
normalized by the relatively bright nitrogen emission.
At higher densities / later times (right panel of \rfig{o_ew_evol}),
when the reverse shock takes over the inner layers of the ejecta, the
oxygen emissivity of the ``subch'' and ``deldet'' models decreases
more rapidly compared to the ``merger''. This is a consequence of the
different oxygen distribution in the inner layers of the explosion
models.

\newcommand\figrow[3]{
        \includegraphics[#1]{ISM#2_#3_rfs_evol}
        \includegraphics[#1]{ISM#2_#3_ratio_sum_rfs_EW_evol}
        \includegraphics[#1]{ISM#2_#3_ratio_sum_rfs_Imax_evol}\\
}
\def\FIGrad [#1] #2 \par{%
  \begin{figure*}\begin{center}%
  	\figrow{#1}{03}{00}
  	\figrow{#1}{03}{70}
  	\figrow{#1}{10}{00}
  	\figrow{#1}{20}{00}
        \caption{#2}
        \label{fig:o_rad}
    \end{center}\end{figure*}
}

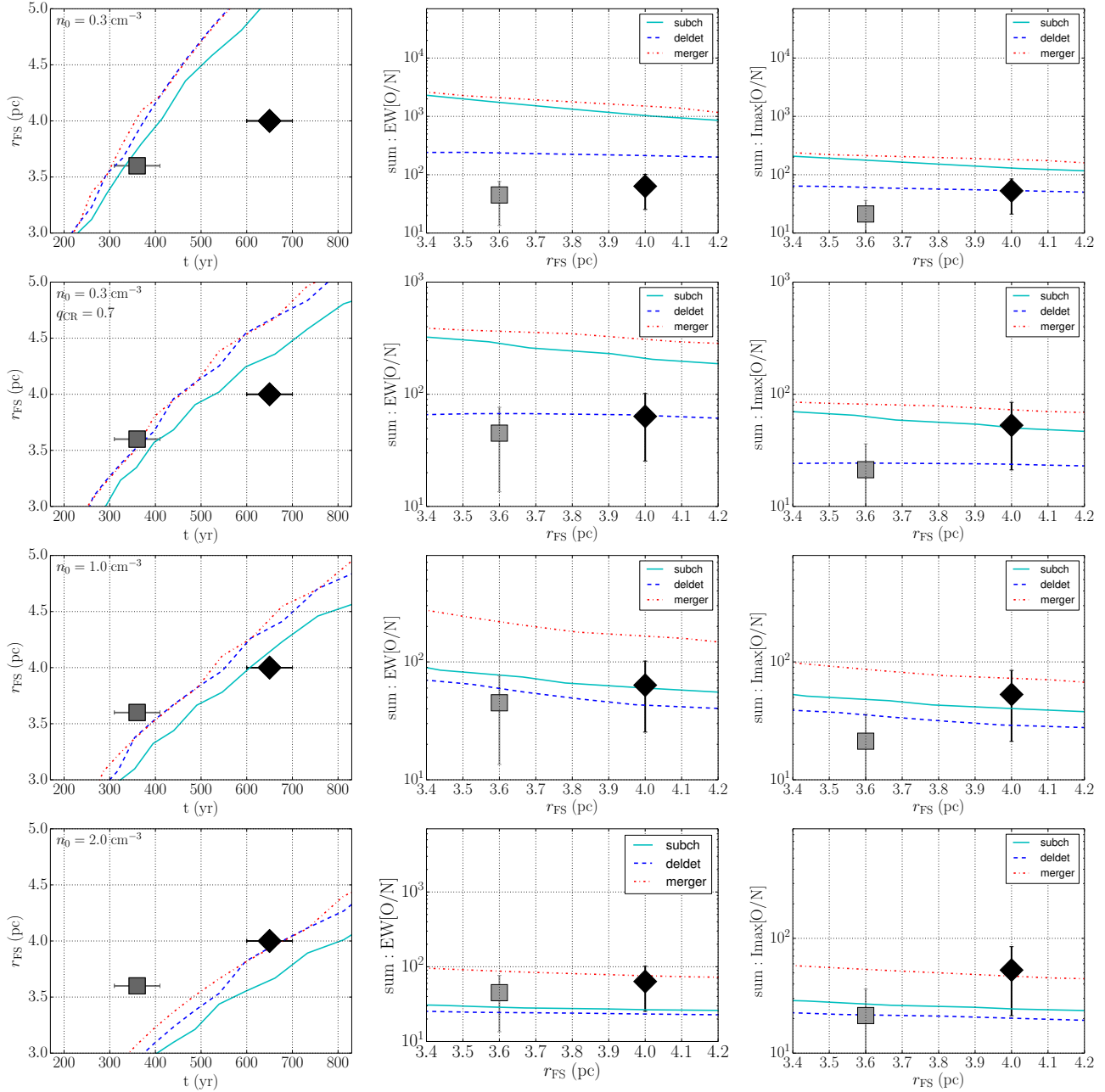
\FIGrad [trim=0 0 40 30,clip=true,width=0.32\hsize,angle=0] Left
column: FS radii as a function of time for different explosion models
(as indicated by the legends in the middle and right columns). 
Middle/right column: evolution of the oxygen to nitrogen EW and flux
ratios, respectively. The evolution of the different models is mapped
against the corresponding FS radius. Squared markers correspond to the
measurements of \Syng\ and diamonds to \Sold. Horizontal error bars are the age 
uncertainty as estimated in \citet{kosenko14}. 
Vertical error bars are derived from the data $1\sigma$ uncertainty of nitrogen flux/EW measurements.
The different rows correspond to different ISM densities as indicated in the labels in
the left column. The simulations in the second row account for
diffusive shock acceleration by cosmic rays (acceleration parameter
$\qcr=0.7$).

A few HD evolutionary models taking into account diffusive shock
acceleration were also created. The modified models were built for
$\nism=0.3\,\mt{cm}^{-3}$ and an acceleration parameter $\qcr\equiv \Pcr/P_\mt{tot}=0.7$, where $\Pcr$ and $P_\mt{tot}$ are the
cosmic ray and the total pressure at the FS, respectively
\citep[developed by][]{kosenko11}. For demonstration purposes this parameter was
chosen higher than typical expected values of $\qcr$ for SNRs; e.g. for Tycho SNR \citet{kosenko11}
found $\qcr\simeq0.5 - 0.7$ (in the assumption of constant cosmic ray diffusion coefficient).

\section{Oxygen in supernova remnants in the Large Magellanic Cloud}
\label{lmcdata}

To determine whether the difference in the oxygen emission line
properties can in principle be detected in SNR
spectra, we considered \emph{XMM-Newton} observations of two young
remnants in the LMC: \Syng\ (OBSID 0111130201) and \Sold\ (OBSID
0113000501). We analyzed their RGS \citep{denherder01} data using
the standard method described in the \emph{XMM-Newton} ABC Guide
(cleaned for soft protons and subtracted the background).

The spectra in the range of
($0.45-0.7$)\,keV were fitted with a power law and four Gaussians for the four lines of
interest (\ions{C}{vi}, \ions{N}{vii}, \ions{O}{vii}, and
\ions{O}{viii} with the {\sc xspec}\footnote{Version 12.7.1} software
package \citep{arnaud96}. The flux ratios were calculated as the
ratios of the respective normalizations, the EW values are provided by
the {\sc eqwidth} utility applied to each Gaussian component.

Due to the poor data statistics, we did not apply the Galactic foreground absorptions in these fits. 
This simplification is acceptable, because we do not use absolute values of fluxes. 
The best fit photon index for \Syng\ is $\sim 3.8\pm0.4$  and for \Sold\ is $0.7\pm 0.2$. 
These measurements are in agreement with the low column density ($n_\mt{H} < $\ee{21}$\mt{cm}^{-2}$) found in \citet{warren04} in the \Syng\ direction. In addition, spectral fitting of \citet{kosenko08} shows no evidence of the Galactic interstellar absorption for this SNR.  For \Sold, however, \citet{kosenko10} reported  $n_\mt{H} \simeq 2.6$\e{21}$\mt{cm}^{-2}$.

Even though the $\chi^2/$d.o.f. values of the fits are satisfactory
(0.99 for \Syng\ and 0.98 for \Sold), the quality of the data does not
allow to estimate the parameters of the lines with adequate precision.
The relative errors of the weak nitrogen line are rather large,
reaching $\sim70\%$ ($1 \sigma$). Thus, the derived fluxes and EWs
should be used with caution and may serve only as approximate
indications. 

In order to confront the data with the models it is essential to know
the dynamical properties and evolutionary stages of the remnants with
sufficient accuracy. The analysis of \Syng\ data reported by
\citet{kosenko08} gives an ISM density of $\nism = (0.4-0.6)\cc$,
\citet{warren04} measured $\nism = 0.05\cc$, and an SNR parametric
analysis reported in \citet{kosenko14} constrains $\nism$ 
to $(0.1-0.3)\cc$ (for an SN explosion of \ee{51}erg). Using
light echo measurements \citet{rest05} limited the age of \Syng\
to $400\pm120$\,yr. Measurements of \cite{ghavamian07} give an age in
the range $300-600$\,yr. 
\citet{badenes08} suggested that the remnant originated 400 years ago
and expands through an ISM with $0.4\cc$. They confirmed the results
of \citet{rest08} where it was found that the SNR originated from an explosion 
similar to SN 1991T, releasing an energy of $1.4$\e{51}erg and producing 
almost one solar mass of $^{56}$Ni.

The \emph{XMM Newton} X-ray data from \Sold\ studied by
\citet{kosenko10} provides an ISM density estimate of $(2.4\pm0.2)\cc$
and $\sim0.4\msun$ of oxygen in the shocked ejecta. Note, that the
oxygen abundance derived from the spectral fitting is about four times
higher than in \Syng. The parametric studies \citep{kosenko14} give $\nism =
(0.5-1.0)\cc$ (for an explosion energy of \ee{51}erg) and the light echo measurements of
\citet{rest05} indicate an age of $600\pm200$\,yr.

Even though in \citet{kosenko14} the ages for \Syng\ and \Sold\ were constrained
to $310-410$\,yr and $600-700$\,yr, respectively, these spans are still too
wide to identify the dynamical stages of these remnants. On the other
hand, their angular dimensions are measured with relatively high
accuracy due to the known distance to the LMC. \cite{warren04}
reported a radius of 3.6\,pc for the shell of \Syng\ and the radius of
the FS in \Sold\ is 4.0\,pc \citep{kosenko10}. Thus, in
\rfig{fig:o_rad} we show the evolution of the
[\ions{O}{vii}+\ions{O}{viii}]/[\ions{N}{vii}] ratios of the EWs and
fluxes mapped against the corresponding radius instead of the
remnant's age (the middle and the right columns). The HD SNR evolutionary
models\footnote{The ISM metallicity is 0.3 of solar. } numerically computed by {\sc supremna} for ISM densities of
$n_0=0.3\cc$, $n_0=0.3\cc$ with DSA, $n_0=1.0\cc$, and
$n_0=2.0\cc$ are depicted in the left columns of
\rfig{fig:o_rad}.  The squares and the diamonds correspond to the measurements of \Syng\
and \Sold\ correspondingly.

The dynamical evolution profiles (left column) and the
data show that for {\bf \Syng} the acceptable ISM density lies
approximately in the range of $(0.2-0.7)\cc$ for the ``subch'' 
model 
and of $(0.3-1.0)\cc$ for the ``deldet'' and ``merger'' models. 
This implies that for these ranges of the ISM density in our set up the dynamic of \Syng\ matches the dynamic of the corresponding models. The range for the ``subch'' model is not explored by \rfig{fig:o_rad} and can be estimated from either linear interpolation between $0.3\cc - 330$ yr and $1.0\cc - 470$ yr or adopting locally Sedov ($\nism \propto t^2$) expansion in the vicinity of $0.3\cc \propto 330$ yr and $1.0\cc \propto 470$ yr.

On the other hand the oxygen emission profiles (in the middle and the right
columns) exclude our ``subch'' model for $\nism\lesssim0.5$ (for
simplicity no DSA assumed). In contrast, our ``deldet'' model matches
the oxygen emissivity of \Syng\ in the required dynamics range of
$\nism = (0.3-1.0)\cc$. However, keep in mind, that CR acceleration
will change these estimates towards slightly lower values of $\nism$.

For {\bf \Sold} the dynamical evolution requires
$\nism\simeq(0.8-1.5)\cc$ for the ``subch'' model, and
$\nism\simeq(1.5-2.5)\cc$ for the ``deldet'' and ``merger''
models. Even though, the available precision of the oxygen emission
data does not exclude the ``subch'' model for this remnant, the
measured amount of shocked oxygen of $\sim0.4\msun$ \citep{kosenko10}
renders this case implausible. According to these plots, our ``merger''
model matches the oxygen emissivity of \Sold\ for the ISM density
range of $(1.5-2.5)\cc$. However, at this stage the quality of the
data (and the statistical significance of the fit) does not allow to
make a definitive prediction.

\section{Results and discussion}
\label{results}

The method we describe here allows to constrain the explosion
mechanism and progenitor scenario of thermonuclear SNe by
investigating emission properties of their remnants. It is based on
modeling the soft X-ray spectra produced by the shocked SN ejecta,
which can be confronted with the observations of SNRs.
This technique can be applied to any remnant of a thermonuclear
explosion in order to constrain the nature of its explosion.
Comparison of the model spectra with high spectral resolution RGS
spectra from two LMC SNRs suggests that a robust identification of the
progenitor scenario may be achieved with high quality data.

Note that for our simulations of the remnant phase we had to use
spherically averaged versions of the 3D explosion models described in
\sect{models}, which are generically asymmetric. This may affect the
shape and strength of the predicted line features. A proper modeling
of the synthetic spectra requires a 3D treatment of the SNR evolution
which could be addressed in future studies.

By relating the oxygen emission lines to those of nitrogen we grade
most of the uncertainties in the nature of the underlying continuum.
The corresponding emission lines lie close to each other in the
spectrum and originate from adjacent regions in the shell, however
separated by the contact discontinuity.
Estimates of the oxygen mass in the remnants of \Syng\ and \Sold\ reported in \citet{kosenko08} and \citet{kosenko10} are subject to many uncertainties that stem from the poorly known conditions around the SNRs. The robustness of these results should be considered with caution (see the relevant discussions in the aforementioned studies) and they require additional confirmation from other sources and methods. The approach presented here combined with high quality data has the potential to reconstruct basic properties of the explosion mechanism. The evolutionary plots can serve as a tool for a prompt classification of any SNR with the available soft X-ray data. Observational data points can be straightforwardly derived from the spectra. No complex and cumbersome multicomponent spectral modelling is required.

Our best guess for the ISM density of $\sim0.5\cc$ for \Syng\ agrees 
with almost all other measurement and with HD models where an explosion of $1.4\times$\ee{51}erg is adopted.
Using the presented technique we find that our delayed detonation model ($1.4\times$\ee{51}erg, $\sim0.6\,M_\odot$ of $^{56}$Ni) agrees better with the data than the merger for \Syng. However, neither of them is a good model for a 91T-like explosion
as inferred for this remnant \citet{rest08}  and \citet{badenes08}. The latter is more abundant than ``deldet'' in $^{56}$Ni ($\sim$1\Msun), 
but of the same explosion energy. If that is the case, our comparison  with the ``subch'' and the ``deldet'' models may be significantly affected. 
For these two models, the amount of oxygen in the ejecta will be significantly lower if the $^{56}$Ni mass increases to 1\Msun.
Therefore, it can be seen as an indication in which direction to search for a good model for this particular remnant at best. Once a detailed simulation of SN1991T explosion becomes available it can be applied to \Syng\ in this framework.

It is still unclear how the explosion mechanism affects the morphology of SNRs expanding into 
a uniform medium. \Syng\ shows a rather spherical shape \citep[Fig. 1 in ][]{warren04} which seems 
natural for single degenerate scenario. For merger remnants  in contrast one would expect a more 
asymmetrical structure. However, a study of \citet{schaefer12} showed no evidence of a surviving companion star
in \Syng, challenging an origin from a delayed detonation
in a single-degenerate progenitor and indicating a WD merger. As this remnant does 
not show an oxygen overabundance in its spectra, this could imply an oxygen-poor merger 
scenario, which could occur if the secondary star is not disrupted, leaving ejecta much 
more similar to a ``subch'' model than our ``merger''.  However, in this case there will also be a compact
remnant (the secondary WD) left, although very faint and probably undetectable.

In contrast, \Sold\ has a tilted axisymmetric morphology
\citep[Fig. 1 in][]{kosenko10} and our analysis suggests that the most
plausible progenitor for this remnant is an oxygen-rich merger. This
is in concordance with the high oxygen abundance found by
\citet{kosenko10} and also in agreement with conclusions of
\citet{edwards12} derived from the absence of a companion star in
the center of the SNR. In this scenario, the respective ISM density of
$\sim2.0\cc$ reconciles the X-ray measurements and the HD simulations
with explosion energy of $1.6\times$\ee{51}erg.
Nevertheless, considerably higher quality of the RGS data is required
in order to derive more reliable conclusions on these SNRs.

We note that a SN type Ia progenitor can modify the ambient medium
through mass outflow and produce a structured and probably clumpy
bubble. This should have noticeable effect on the remnant evolution.
Moreover, a wind from a pre-supernova can modify the chemical 
composition of the circumstellar medium (CSM), which in turn could affect 
nitrogen abundance and thus its emissivity. \citet{chiotellis13} have shown how a non-uniform 
CSM around a SN progenitor will affect the SNR's evolution and
eventually also its X-ray emission. They modelled the remnant of SN
1572 (Tycho) by means of the {\sc supremna} code. 
They also analysed the impact on the X-ray emission from 
chemically enriched AGB winds, and found that the changes in the integral thermal emission 
are not noticeable. It is conceivable that in some special cases the 
nitrogen abundance in the CSM could be modified by the progenitor, thus slightly altering our criteria,
however the current accuracy of the measurements would not allow to 
distinguish such a minuscule difference.

We note that within the scope of the present paper we focus on the variations caused by
different explosion mechanisms only. A more detailed approach would
require self-consistent models of stellar evolution and an explosion
model coupled to a realistic wind profile. The presence of dense
clouds in the vicinity of a supernova could alter the overall
properties the remnant's X-ray emission, but in comparison to the
effects caused by differences in the chemical composition of the
ejecta these effects should be small. In addition, we note that it is
commonly assumed that these LMC remnants expand into a uniform
medium.



\section{Conclusion}
\label{conclusion}

A method based on state-of-the-art 3D simulations of thermonuclear SN
explosions, coupled with hydrodynamic calculations of the SNR's
evolution, making use of the most up-to-date atomic data, was used to
study the properties of oxygen emission lines in synthetic X-ray
spectra. It was shown that confronting these models with observations
of young SNRs can help to identify the nature of their
progenitors. Our results suggest that for \Sold\ an oxygen-rich merger
is the preferred scenario whereas \Syng\ could be either the result of
a delayed-detonation explosion or an oxygen-poor merger. Even though, at
present, the quality of data available is insufficient to
draw robust conclusions, combining different observations and
data allows narrowing down possible scenarios.

\vskip 1cm \textit{Acknowledgments.} 
We thank the anonymous referee for useful discussions and 
comments which considerably helped to improve the manuscript. 
DK was supported by RFBR
grant 13-02-92119 and by the French national research agency ANR (COSMIS project). 
She thanks the Max-Planck-Institut f\"ur Astrophysik for the warm hospitality. SB was supported by RFBR grant 13-02-92119 and SAI MSU. This work was supported by the
Deutsche Forschungs\-gemeinschaft via the Transregional Collaborative
Research Center TRR 33 `The Dark Universe' and the Excellence Cluster
EXC153 `Origin and Structure of the Universe'.

\setlength{\bibhang}{2.0em}
\setlength\labelwidth{0.0em}

\bibliographystyle{mn2e}
\bibliography{supremnamodels}

\end{document}